\documentclass[final,3p,authoryear]{elsarticle}
\usepackage{amssymb}

\journal{Icarus}

\begin{document}

\begin{frontmatter}
\title{The Nucleus of Main-Belt Comet P/2010 R2 (La Sagra)
}
\author[asiaa,uh]{Henry H.\ Hsieh}\ead{hhsieh@asiaa.sinica.edu.tw}
\address[asiaa]{Institute of Astronomy and Astrophysics, Academia Sinica, P.O.\ Box 23-141, Taipei 10617, Taiwan}
\address[uh]{Institute for Astronomy, University of Hawaii, 2680 Woodlawn Drive, Honolulu, Hawaii 96822, USA}

\begin{abstract}
We present recent observations of main-belt comet P/2010 R2 (La Sagra) obtained using the Gemini North telescope on five nights in late 2011 and early 2013 during portions of the object's orbit when dust emission was expected to be minimal or absent.  We find that P/La Sagra continues to exhibit a faint dust trail aligned with its orbit plane as late as 2011 December 31, while no activity is observed by the time of our next observations on 2013 March 3, shortly before aphelion.  Using only photometry measured when the comet was observed to be inactive, we find best-fit IAU phase function parameters of $H_R=18.4\pm0.2$~mag and $G=0.17\pm0.10$, corresponding to an effective nucleus radius of $r_N=0.55\pm0.05$~km (assuming $p_R=0.05$). We revisit photometry obtained when P/La Sagra was observed to be active in 2010 using our revised determination of the object's nucleus size, finding a peak dust-to-nucleus mass ratio of $M_d/M_N = (5.8\pm1.6)\times10^{-4}$, corresponding to an estimated total peak dust mass of $M_d = (5.3\pm1.5)\times10^8$~kg.  We also compute the inferred peak total active surface area and active surface fraction for P/La Sagra, finding $A_{\rm act}\sim5\times10^4$~m$^2$ and $f_{\rm act} \sim0.01$, respectively. Finally, we discuss P/La Sagra's upcoming perihelion passage, particularly focusing on the available opportunities to conduct follow-up observations in order to search for recurrent activity and, if recurrent activity is present, to search for changes in P/La Sagra's activity strength on successive orbit passages that should provide insights into the evolution of MBC activity over time. [{\it Accepted by Icarus, 2014 August 21}]
\end{abstract}


\end{frontmatter}

\section{INTRODUCTION}
\label{section:introduction}

\subsection{Background}
\label{section:background}

Main-belt comets \citep[MBCs;][]{hsi06} are objects that exhibit cometary activity as a result of the sublimation of volatile ice, yet occupy stable orbits in the main asteroid belt.  They have attracted interest, particularly in astrobiology, due to the possibility that icy material from the main belt region could have been a significant primordial source of terrestrial water \citep[e.g.,][]{mor00,ray04,obr06}, and their potential as compositional probes of the early inner solar system in general \citep[e.g.,][]{hag09}.  MBCs comprise a subset of the group of objects known as active asteroids, which are objects with asteroidal orbits (defined as having Tisserand parameter values of $T_J>3$) that exhibit comet-like activity, typically in the form of dust emission \citep[e.g.][]{jew12}.  The classification of active asteroids also includes disrupted asteroids \citep[e.g.,][]{jew10,jew11b,jew13c,sno10,bod11,ish11b,ste12b,mor11a,mor12}, which are otherwise inert objects that exhibit comet-like mass loss as a result of collisional or rotational disruption \citep[cf.][]{hsi12a}.

In general, direct confirmation of MBC sublimation via spectroscopic detections of gas emission is beyond the capabilities of current facilities due to the extremely weak sublimation rates of these objects.  Practically speaking, this inability to directly detect sublimation means that none of the objects currently considered to be MBCs can be definitively confirmed as cometary bodies.  However, \citet{hsi12a} outlined a range of methods that can be used to infer the most likely cause of observed dust emission, and therefore distinguish likely MBC candidates (hereafter referred to simply as ``MBCs'') and disrupted asteroid candidates (hereafter, ``disrupted asteroids'').  These methods include monitoring to search for persistent or increasing activity (although this condition alone is insufficient proof of sublimation-driven activity without other supporting evidence), deep high-resolution imaging to search for unusual morphology that could indicate the action of non-cometary causes of activity, numerical dust modeling, and long-term monitoring to search for repeated activity over multiple orbits, which is naturally explained as a result of sublimation-driven activity, but would be highly implausible for activity generated via other mechanisms.

Detection of recurrent activity is currently considered to be the most reliable method of indirectly identifying sublimation-driven activity \citep[cf.][]{jew12}. However, making such a detection requires  waiting for at least an entire orbit period ($\sim$5-6 years for most of the currently known MBCs), and potentially more depending on observational circumstances, to elapse following the discovery of activity in a new MBC before repetition of activity can be confirmed or ruled out.  As such, recurrent activity has only been successfully confirmed to date for MBCs 133P/Elst-Pizarro and 238P/Read \citep{hsi04,hsi10b,hsi11b,hsi13b}, although unsuccessful attempts to detect recurrent activity for MBC 176P/LINEAR have also been made \citep{dev12,hsi14}.

While even recurrent activity does not ensure that any observed activity is actually sublimation-driven, the recent spectroscopic confirmation of water vapor outgassing on (1) Ceres by the {\it Herschel Space Observatory} \citep{kup14} has unequivocally demonstrated that currently sublimating icy material is in fact present in the main asteroid belt.  This water vapor detection strongly supports the physical plausibility of sublimation-driven cometary activity on other main-belt objects, such as the MBCs.  Additionally, spectroscopic features attributed to water ice frost have been reported for main-belt asteroid (24) Themis \citep{riv10,cam10} and outer main-belt asteroid (65) Cybele \citep{lic11,tak12}.  These results are particularly interesting in the context of this work because MBCs 133P, 176P, 288P, and possibly 238P \citep{hag09,hsi12b} have all been dynamically linked to the Themis asteroid family, of which Themis is the largest member.  Despite these recent discoveries of corroborating evidence of water ice in the asteroid belt, however, it is important to note that Ceres \citep[952 km in diameter][]{tho05}, Themis \citep[202 km in diameter;][]{wri10,mai11,mas11} and Cybele \citep[273 km in diameter;][]{mue04} are all much larger than the largest known MBCs (cf.\ Section~\ref{section:mbc_comparison}) and therefore likely differ significantly from the MBCs in both their thermal evolutionary history and present-day geophysical nature \citep[e.g.,][]{fan89,mcc11}.  As such, any comparisons between such large asteroids and the km-scale MBCs must be regarded with caution.

\subsection{P/2010 R2 (La Sagra)}
\label{section:intro_lasagra}

The fifth MBC to be discovered was Comet P/2010 R2 (La Sagra), which was found on 2010 September 14.9 (UT) by the 0.45~m telescope at La Sagra Observatory in southern Spain \citep{nom10}.  Numerical modeling of the object's dust emission indicated that the observed mass loss was best simulated by anistropic comet-like dust emission at a level of $\sim$3-4~kg~s$^{-1}$ persisting for at least 7 months following perihelion, with the apparent source of emission located near the south pole of the nucleus and the object having an obliquity near 90$^{\circ}$ \citep{mor11b}.  Additional analysis of precovery and follow-up observations revealed an increase of $>$1~mag in the comet's brightness between 2010 August and December, indicating the presence of ongoing dust production, consistent with sublimation-driven activity and inconsistent with impulsive activity generation mechanisms such as an impact \citep{hsi12c}.

Spectroscopic observations produced no evidence of gas emission, but were still used to place an upper limit on the comet's CN production rate of $Q_{\rm CN}=6\times10^{23}$~mol~s$^{-1}$.  Assuming average CN/H$_2$O ratios measured for previously observed comets \citep{ahe95}, this result corresponds to an upper limit H$_2$O production rate of $Q_{\rm H_2O}\sim10^{26}$~mol~s$^{-1}$.  Thermal modeling by \citet{pri09} indicates that all other volatile material except for water ice may become severely depleted over the lifetime of a MBC, however, suggesting that the ratio of CN to H$_2$O in MBCs may be much lower than in classical comets from the outer solar system.  If this is true, then the upper limits on the H$_2$O production rates of P/La Sagra and other MBCs inferred from observations of CN emission \citep[e.g.,][]{jew09,jew12,jew14a,hsi12b,hsi12c,hsi13a,lic13} may in fact be far larger than reported.

\citet{hsi12c} additionally reported that numerical dynamical simulations of the object indicated that the region in orbital element space occupied by P/La Sagra is largely stable, despite being crossed by the 13:6 mean-motion resonance with Jupiter and the $(3,-2,-1)$ three-body mean-motion resonance with Jupiter and Saturn.  These results imply that the object is likely to be native to its present-day location, and that its physical properties therefore may be similar of those of other objects in that region of the asteroid belt.  However, given the moderately chaotic nature of the region (due to the presence of nearby resonances) and the possibility of the influence of non-gravitational forces (which were not included in the dynamical analysis), the possibility that P/La Sagra could have formed elsewhere and recently dynamically evolved onto its current orbit could not be definitively ruled out. In fact, preliminary results from dynamical simulations including the Yarkovsky effect \citep{bur79,far99} indicate that Yarkovsky-induced drift of the semimajor axis of P/La Sagra (estimated to be on the order of $3\times10^{-4}$~AU~Myr$^{-1}$) could indeed affect its long-term dynamical stability \citep{nov12}.  Meanwhile, non-gravitational accelerations as large as $2.5\times10^{-1}$~m~s$^{-2}$ have been inferred for 133P based on astrometric measurements \citep{che10}, though the effect of non-gravitational accelerations of this magnitude on the dynamical stability of MBCs has not yet been well-studied.

The physical characterization of the nuclei of individual MBCs and MBC candidates that have been previously observed to be active is extremely important for improving our understanding of how the active MBC population relates to the inactive background population.  Physical studies of MBC nuclei indicate what characteristics dormant MBCs might have, allowing us to better estimate the size of the population of inactive but icy asteroids, constrain thermal modeling studies of MBCs aimed at better understanding volatile preservation in small inner solar system bodies \citep[e.g.,][]{sch08,pri09,cap12}, and enable quantitative analyses of total dust productivity and photometric searches for low-level activity \citep[e.g.,][]{hsi12c,hsi14}.  Physical characterization of MBC activity is also important for improving our understanding of other characteristics of the population, such as peak activity levels, typical duration of activity, and relative activity strength levels from one orbit to the next, that could ultimately shed light on key issues such as potential mechanisms for producing and sustaining activity, the degree of variation in volatile content within the MBC population, and prospects for using MBCs as tracers of the abundance and distribution of ice in the present-day asteroid belt.  With these issues in mind, we set out in this work to measure the size of P/La Sagra's nucleus and analyze previously reported data obtained when P/La Sagra was active in order to quantitatively characterize its activity strength and compare it to other MBCs.

\section{OBSERVATIONS}
\label{section:observations}

We observed P/La Sagra using the 8.1~m Gemini North telescope (Programs GN-2011B-Q-17 and GN-2013A-Q-102) on Mauna Kea in Hawaii.  Observations were made using the imaging mode of the Gemini Multi-Object Spectrograph \citep[GMOS; image scale of $0.1454''$~pixel$^{-1}$;][]{hoo04} and a Sloan Digital Sky Survey (SDSS) $r'$ filter.  Non-sidereal tracking at the apparent rate and direction of motion of P/La Sagra on the sky was used for all observations.

\setlength{\tabcolsep}{5.0pt}
\begin{table}[ht]
\caption{Gemini observations of P/2010 R2 (La Sagra)}
\smallskip
\footnotesize
\begin{tabular}{lccccccccc}
\hline\hline
\multicolumn{1}{c}{UT Date}
 & \multicolumn{1}{c}{N$^a$}
 & \multicolumn{1}{c}{t$^b$}
 & \multicolumn{1}{c}{Filter}
 & \multicolumn{1}{c}{$\nu$$^c$}
 & \multicolumn{1}{c}{$R$$^d$}
 & \multicolumn{1}{c}{$\Delta$$^e$}
 & \multicolumn{1}{c}{$\alpha$$^f$}
 & \multicolumn{1}{c}{$m_R(R,\Delta,\alpha)$$^g$}
 & \multicolumn{1}{c}{$m_R(1,1,\alpha)$$^h$} \\
\hline
2010 Jun 25    & \multicolumn{3}{l}{\it Perihelion...} & 0.0 & 2.623 & 2.239 & 22.4 & ... & ... \\
2011 Sep 25    &  9 &  1620 & $r'$ & 100.5 & 3.111 & 2.922 & 18.8 & 23.8$\pm$0.1   & 19.0$\pm$0.1   \\ 
2011 Dec 31    &  9 &  2700 & $r'$ & 117.0 & 3.249 & 2.297 &  5.2 & 23.0$\pm$0.1   & 18.6$\pm$0.1   \\ 
2013 Mar 03    & 10 &  1800 & $r'$ & 178.8 & 3.570 & 2.695 &  8.6 & 23.9$\pm$0.1   & 19.0$\pm$0.1   \\ 
2013 Mar 12    & \multicolumn{3}{l}{\it Aphelion...} & 180.0 & 3.569 & 2.750 & 10.3 & ... & ... \\
2013 Apr 08    & 10 &  1800 & $r'$ & 183.6 & 3.569 & 3.021 & 14.7 & 24.4$\pm$0.1   & 19.2$\pm$0.1   \\ 
2013 Apr 12    & 10 &  1800 & $r'$ & 184.1 & 3.568 & 3.070 & 15.1 & 24.4$\pm$0.1   & 19.2$\pm$0.1   \\ 
2015 Nov 30    & \multicolumn{3}{l}{\it Perihelion...} & 0.0 & 2.620 & 2.946 & 19.3 & ... & ... \\
\hline
\hline
\end{tabular}
\newline {$^a$ Number of exposures}
\newline {$^b$ Total effective exposure time, in seconds}
\newline {$^c$ True anomaly, in degrees}
\newline {$^d$ Heliocentric distance, in AU}
\newline {$^e$ Geocentric distance, in AU}
\newline {$^f$ Solar phase angle (Sun-P/La Sagra-Earth), in degrees}
\newline {$^g$ Mean apparent magnitude}
\newline {$^h$ Mean apparent magnitude, normalized to $R=\Delta=1$~AU}
\label{table:obslog_p2010r2}
\end{table}

\begin{figure}
\centerline{\includegraphics[width=3.0in]{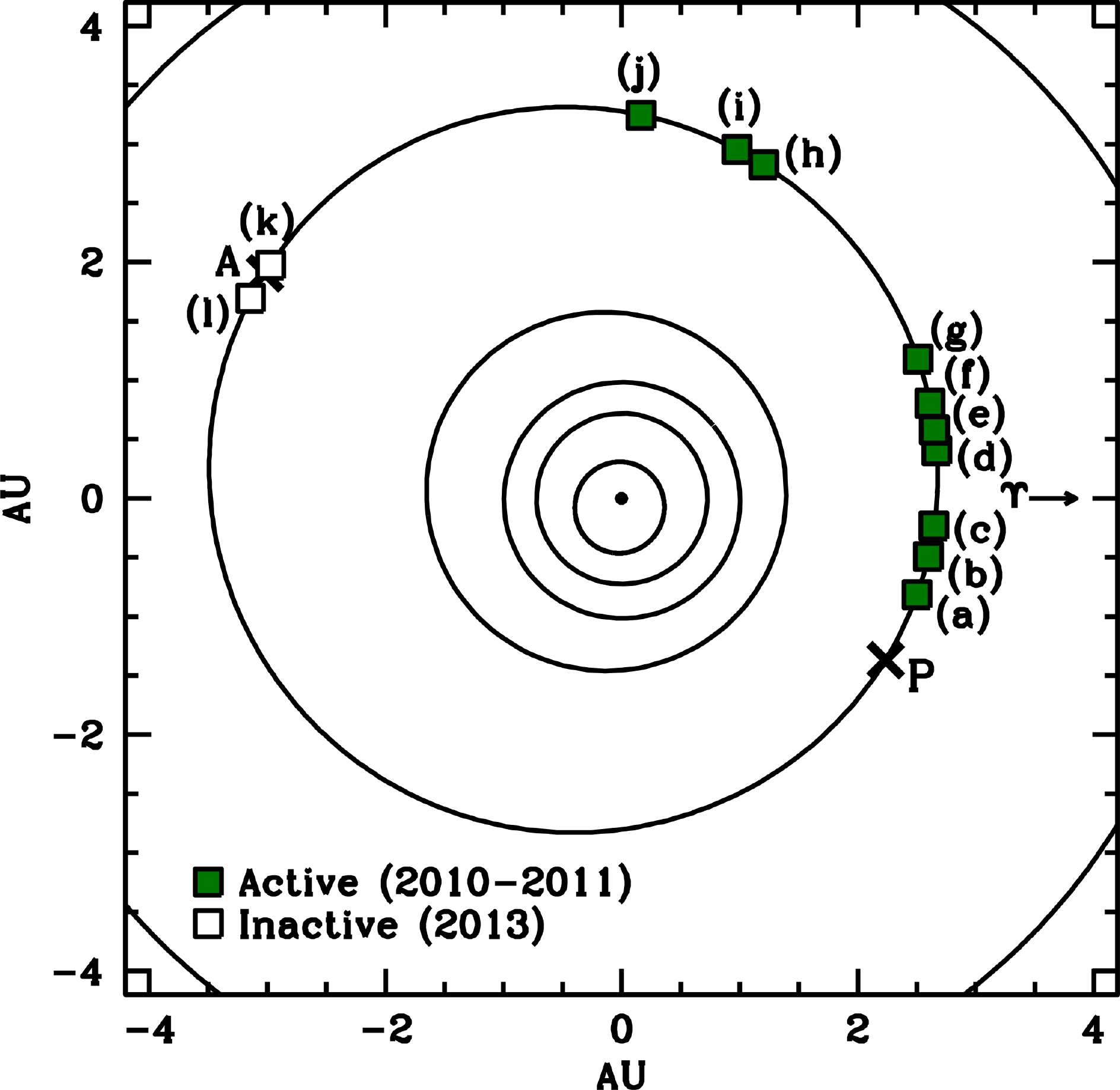}}
\caption{\small Orbital position plot of P/La Sagra observations detailed in Table~\ref{table:obslog_p2010r2}. The Sun is shown at the center as a solid dot, with the orbits of Mercury, Venus, Earth, Mars, P/La Sagra, and Jupiter (from the center of the plot outwards) shown as black lines. Solid squares mark positions where P/La Sagra was observed to be active in 2010 and 2011, while open squares mark the position where P/La Sagra was observed to be inactive in 2013. Perihelion (P) and aphelion (A) positions are marked with crosses. Observations plotted were obtained on (a) 2010 August 16, (b) 2010 September 8-19, (c) 2010 October 5, (d) 2010 November 25-28, (e) 2010 December 12, (f) 2010 December 31, (g) 2011 February 3, and (h) 2011 August 26-31 \citep{hsi12c}, (i) 2011 September 25, (j) 2011 December 31, (k) 2013 March 3, and (l) 2013 April 8-12 (this work).
}
\label{figure:orbitplot}
\end{figure}

\begin{figure}
\centerline{\includegraphics[width=5.0in]{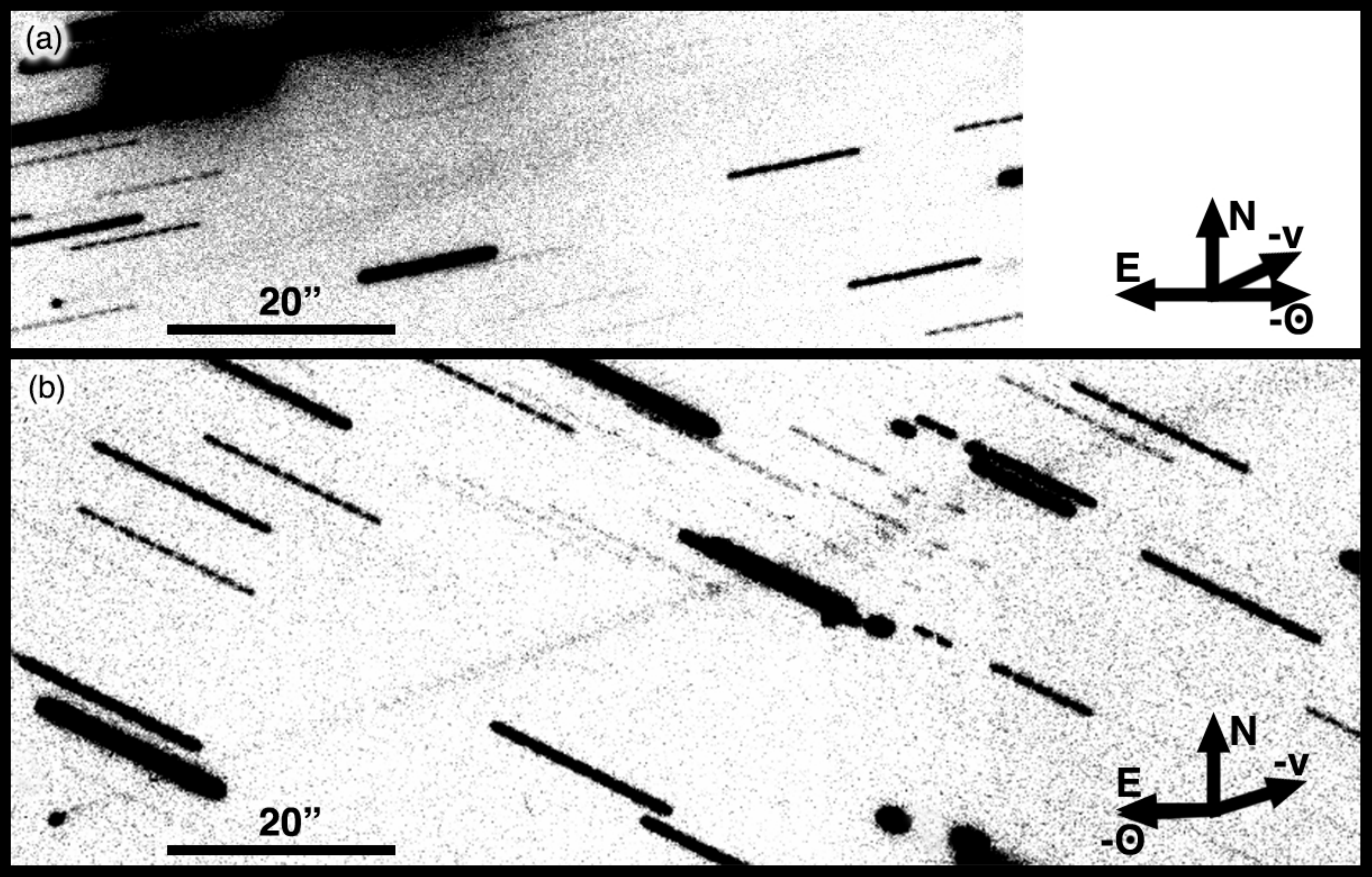}}
\caption{\small Composite images of P/2010 R2 (La Sagra) constructed from data obtained on (a) UT 2011 September 25 (total effective exposure time of 1620~s), and (b) UT 2011 December 31 (total effective exposure time of 2700~s).  The nucleus of the comet is at the extreme lower left corner of each panel with a faint dust trail visible extending towards the upper right corner.  The upper panel is $2.0'\times0.5'$ in angular size, while the lower panel is $2.0'\times0.75'$.  Arrows indicate north (N), east (E), the negative heliocentric velocity vector ($-$v), and the direction toward the Sun ($\odot$).
}
\label{figure:images_2011}
\end{figure}

We obtained observations on two nights in late 2011 and three nights in early 2013 during portions of the object's orbit when dust emission was expected to be minimal or absent.    Details of these observations are shown in Table~\ref{table:obslog_p2010r2}, while orbital positions of all reported photometric measurements are shown in Figure~\ref{figure:orbitplot}.  Images from each night were aligned on the object and summed together to create composite images in order to check for activity.  A long dust trail aligned with the projection of the object's orbit plane in the sky, likely composed of large dust particles remaining from the 2010 active period \citep[e.g.,][]{rea00,rea07,aga10}, is observed in data obtained in 2011 (Figure~\ref{figure:images_2011}), while no activity is apparent in any of our data from 2013 (Figure~\ref{figure:images_2011}).  In all of our data, whether the dust trail is visible or not, measurements of the PSF width of composite images of the comet in the direction perpendicular to its direction of apparent non-sidereal motion on the sky show no deviations from the PSF widths of field stars measured in the same way, indicating the absence of any resolved coma during these observations.

Standard image calibration (bias subtraction and flat-field reduction) was performed for all images.  Flat fields were constructed from dithered images of the twilight sky.  Object and field star photometry was performed using circular apertures for which optimum sizes were chosen accounting for the amount of field-star trailing caused by the non-sidereal tracking of the object, and nightly seeing conditions.  Absolute calibration of object photometry was performed using field star magnitudes from the Pan-STARRS1 survey's photometric catalog \citep[cf.][]{ton12,sch12,mag13}.  Transformation of $r'$-band magnitudes to Kron-Cousins $R$-band magnitudes is performed for all photometric results, assuming solar colors for the object, for more straight-forward comparison to previously measured $R$-band MBC nucleus absolute magnitudes \citep[e.g.,][]{hsi10b,hsi11a,hsi11b,mac12}.

\begin{figure}
\centerline{\includegraphics[width=5.0in]{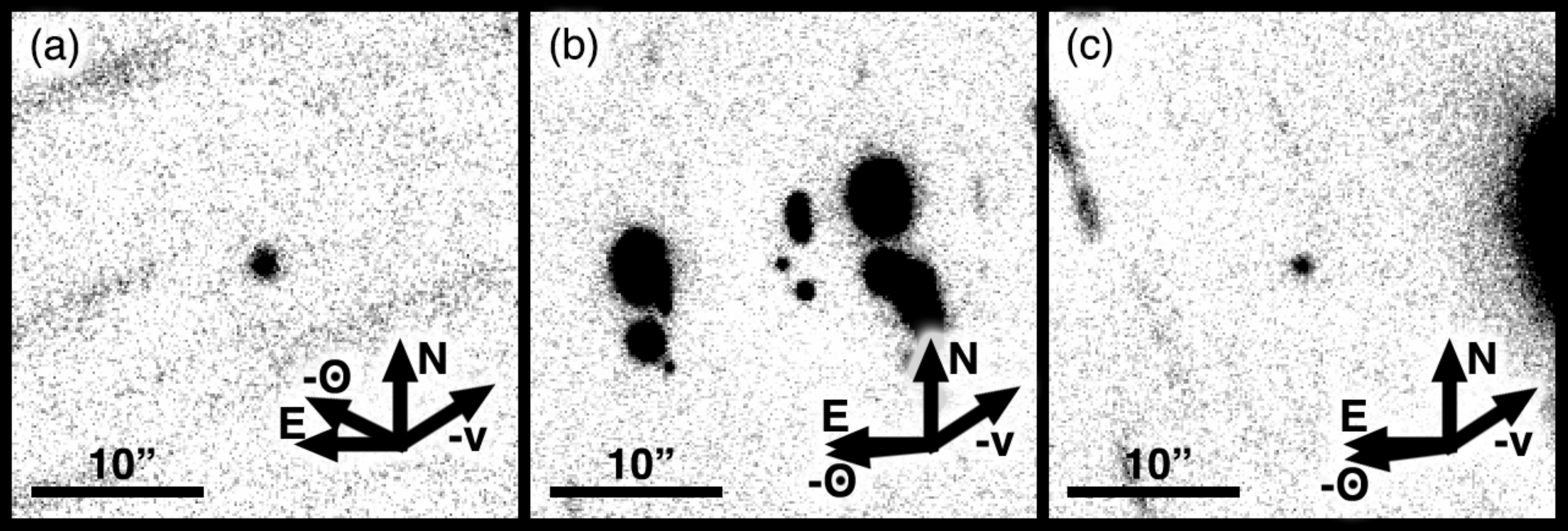}}
\caption{\small Composite images of P/2010 R2 (La Sagra) constructed from data obtained on (a) UT 2013 March 3 (total effective exposure time of 1800~s), (b) UT 2013 April 8 (total effective exposure time of 1800~s), and (c) UT 2013 April 12 (total effective exposure time of 1800~s).  The object is at the center of each panel, where no activity is detected in any of our images.  All panels are $0.5'\times0.5'$ in angular size.  Arrows indicate north (N), east (E), the negative heliocentric velocity vector ($-$v), and the direction toward the Sun ($\odot$).
}
\label{figure:images_2013}
\end{figure}

\section{ANALYSIS AND RESULTS}

\subsection{Phase Function Analysis}
\label{section:phsfn_analysis}

In order to determine the properties of P/La Sagra's nucleus, we only consider data where no visible activity is detected, i.e., only data from 2013.  We prepare the data for the determination of P/La Sagra's phase function by first normalizing the measured apparent magnitudes, $m(R,\Delta,\alpha)$, to unit heliocentric and geocentric distances, $R$ and $\Delta$, respectively (i.e., $R=\Delta=1$~AU), where $\alpha$ is the solar phase angle using
\begin{equation}
m(1,1,\alpha) = m(R,\Delta,\alpha) - 5\log (R\Delta)
\end{equation}
The resulting reduced magnitude, $m(1,1,\alpha)$, remains dependent on the solar phase angle via the solar phase function, as well as the rotational phase of the nucleus at the time of the observations in question.  At this time, the rotational properties of P/La Sagra are unknown, and since all of our observations were short-duration ``snap-shot'' observations, we are unable to constrain any of these rotational properties from the data in hand.  For the purposes of this analysis, however, we simply incorporate the unknown brightness variations of the nucleus due to rotational effects into the overall uncertainty of each of our photometry points.  For the purposes of estimating this uncertainty, we assume a peak-to-trough rotational photometric range for P/La Sagra's nucleus of $\Delta m = 0.6$~mag, where the peak photometric ranges of 133P's nucleus and 176P's nucleus have been measured to be $\Delta m=0.4$~mag and $\Delta m=0.7$~mag \citep{hsi04,hsi10b,hsi11a}, respectively, and results from \citet{mas09} indicate that $\sim$80\% of main-belt asteroids should have photometric ranges of $\Delta m \leq 0.6$~mag.

\begin{figure}
\centerline{\includegraphics[width=3.3in]{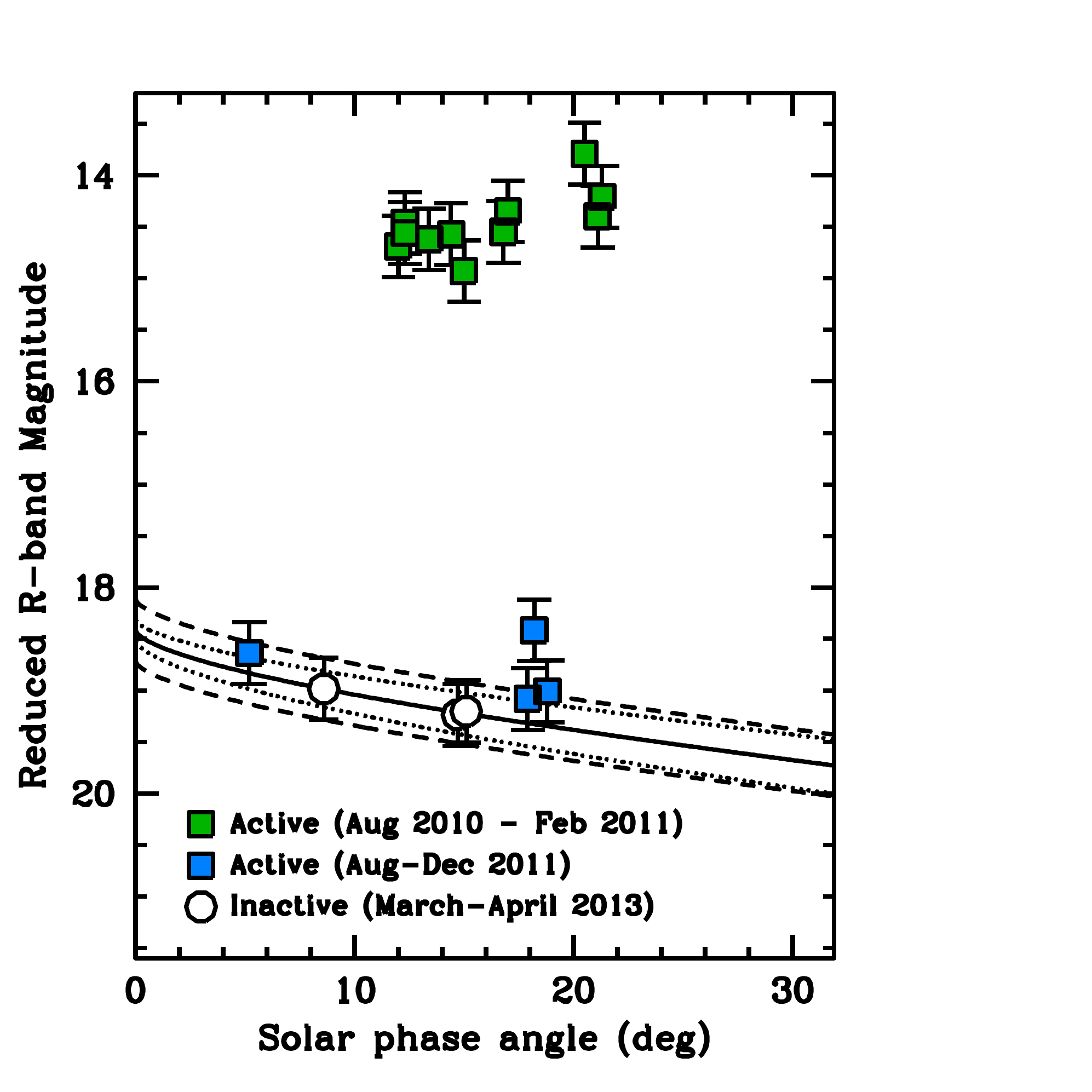}}
\caption{\small Best-fit IAU phase function (solid line) for P/La Sagra, where dotted lines indicate the range of uncertainty due to phase function parameter uncertainties, and dashed lines indicate the expected possible photometric range due to rotational brightness variations, assuming a peak-to-trough photometric range of $\Delta m=0.6$~mag.  Open circles denote photometry measured in this work when the comet was observed to be inactive, while green filled squares mark photometry obtained by \citet{hsi12c} when the comet was extremely active in 2010 and early 2011, and blue filled squares mark photometry obtained in this work and by \citet{hsi12c} when the comet was weakly active in the latter half of 2011.
}
\label{figure:phase_function}
\end{figure}

We find best-fit parameters of $H_R=18.4\pm0.2$~mag and $G_R=0.17\pm0.10$ for P/La Sagra, using the standard IAU $H,G$ formalism \citep{bow89}, where the estimated uncertainties are dominated by the nucleus's unknown rotational properties and its unknown rotational phase at the time of our observations.  There is also some uncertainty associated with the fact that our observations of the inactive nucleus span a limited phase angle range (from $\alpha=8.6^{\circ}$ to $\alpha=15.1^{\circ}$), and so future observations of the inactive nucleus at phase angles outside of this range, particularly at low phase angles (i.e., $\alpha\sim0^{\circ}$) to measure any opposition surge that the object might exhibit \citep[cf.][]{bel00}, will be extremely useful for refining the parameters found in this work.

We plot our best-fit phase function as well as the data used to fit it in Figure~\ref{figure:phase_function}.  We also plot photometric points that were measured for P/La Sagra when it was observed to be active.  The absolute magnitude computed here is fainter that that estimated by \citet{hsi12c} because that previous analysis used photometric data that we now know was contaminated with unresolved near-nucleus dust emission (cf.\ Section~\ref{section:activedata}).  Our computed slope parameter is within the range of previously measured MBC nucleus slope parameters of $G=0.04\pm0.05$ for 133P, $G=0.15\pm0.05$ for 176P, $G=-0.03\pm0.10$ for 238P, $G=-0.08\pm0.05$ for 259P, and $G=0.05\pm0.05$ for P/2006 VW$_{139}$ \citep[][Hsieh et al., in prep]{hsi09b,hsi10b,hsi11a,hsi11b,mac12}, as well as within the range of slope parameters measured for kilometer-scale asteroids in the Themis family (with which 133P, 176P, and P/2006 VW$_{139}$ are associated) \citep{hsi08}.  

\begin{figure}
\centerline{\includegraphics[width=3.5in]{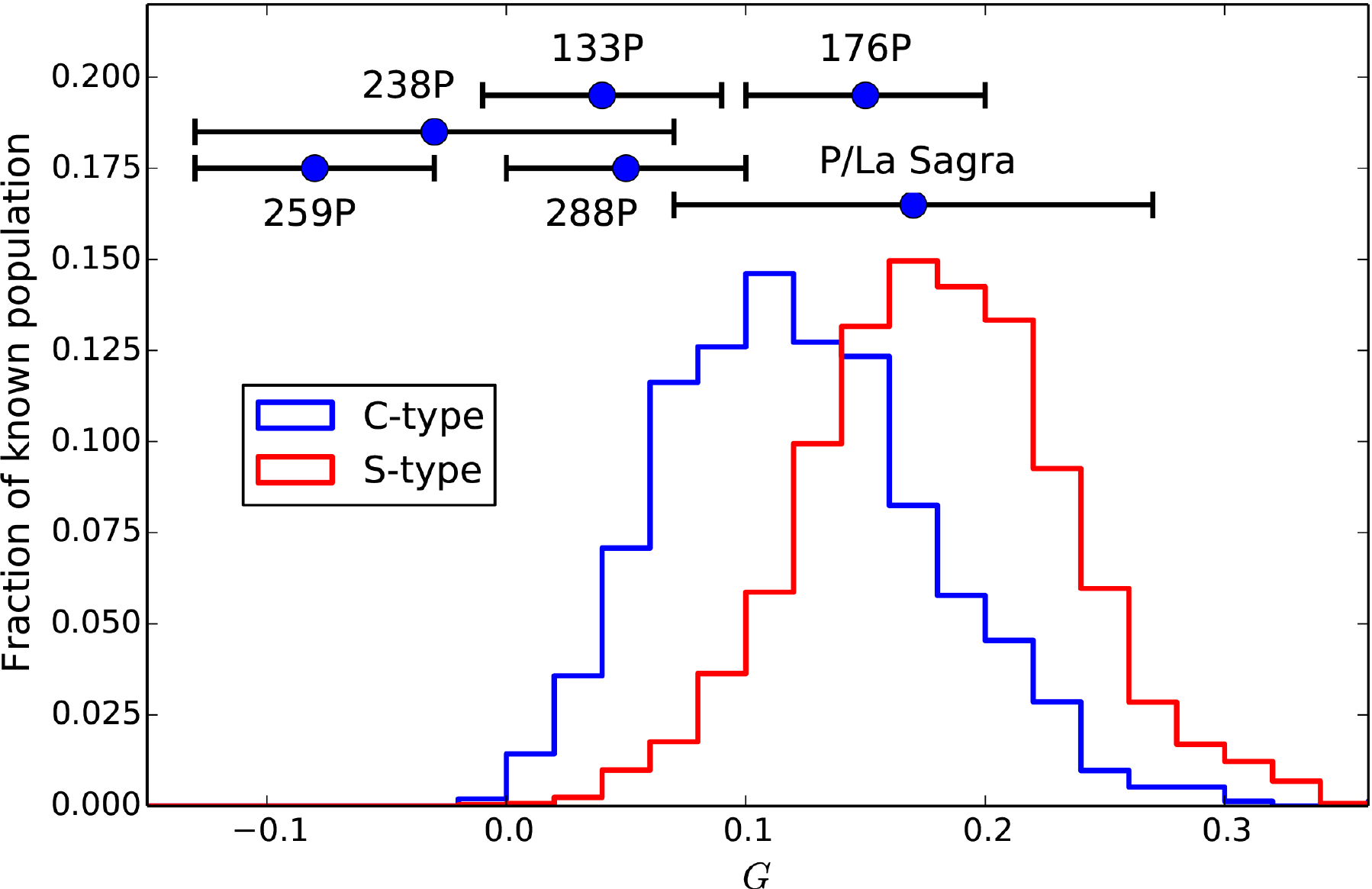}}
\caption{\small Normalized histograms of $G$ parameter values provided online at {\tt https://wiki.helsinki.fi/display/PSR/ Asteroid+absolute+magnitude+and+slope} \citep{mui10,osz11} for C- and S-type asteroids (blue and red lines, respectively) as classified according to SDSS observations \citep{car10,has12}.  For reference, $G$ values and 1-$\sigma$ uncertainities reported for MBCs are also plotted as blue circles.
}
\label{figure:taxonomy_g}
\end{figure}

\begin{figure}
\centerline{\includegraphics[width=2.8in]{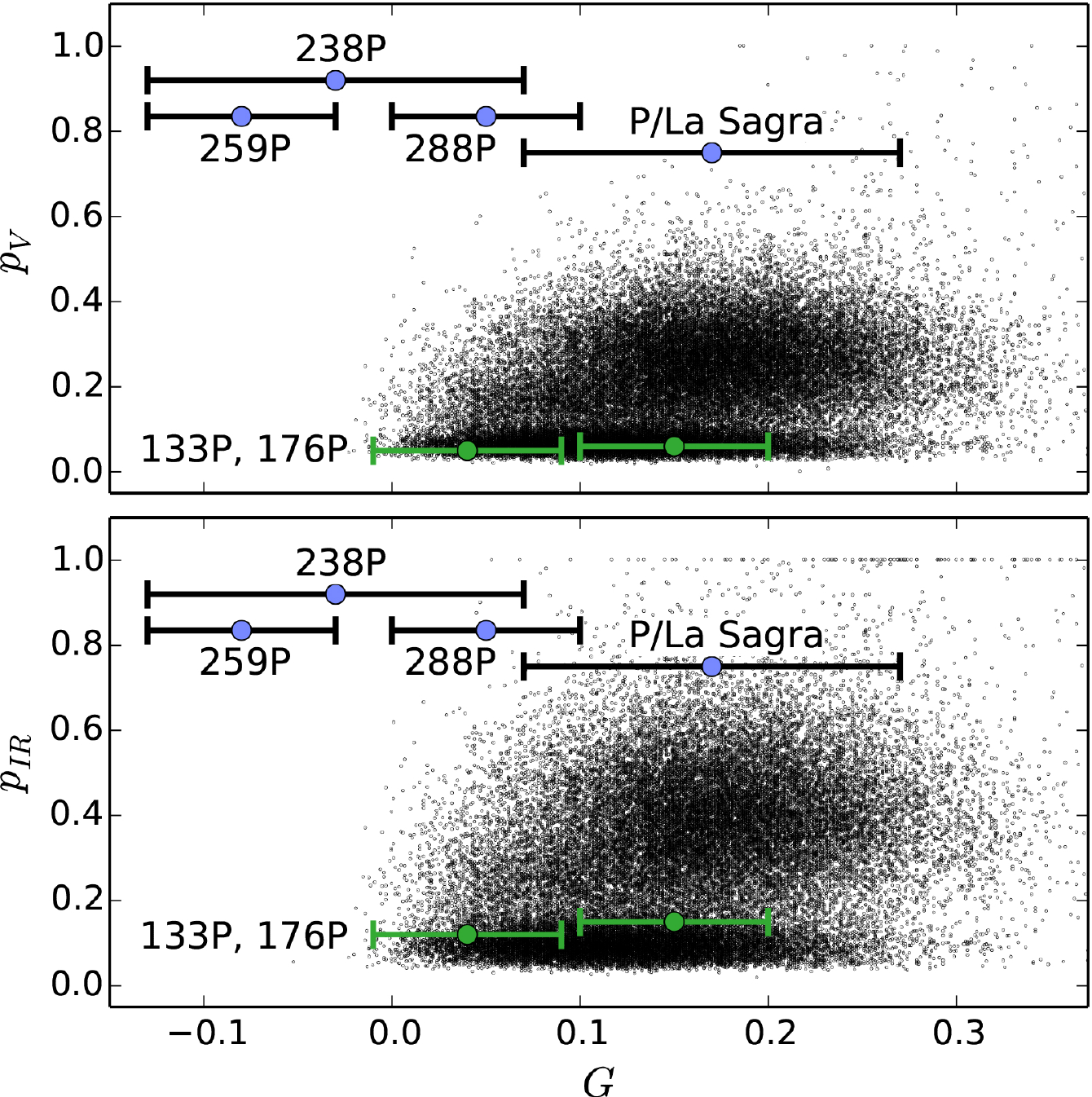}}
\caption{\small Plots of $G$ parameter values \citep[provided online at {\tt https://wiki.helsinki.fi/display/PSR/ Asteroid+absolute+magnitude+and+slope};][]{mui10,osz11} versus visible albedos (top panel) and infrared albedos (lower panel) for numbered asteroids measured by \citet{mas11}.  For reference, we also plot green circles indicating $G$ values, 1-$\sigma$ uncertainties, and visible and infrared albedos reported for 133P and 176P \citep{hsi09b,bau12}, as well as blue circles indicating $G$ values and 1-$\sigma$ uncertainties reported for other MBCs for which albedos are unavailable.  In the cases of 133P and 176P, we assume that $p_R\approx p_V$ \citep[likely to be the case given the objects' largely neutral visible spectra;][]{lic11}, where the 1-$\sigma$ uncertainties in albedo are roughly the size of the symbols plotted.
}
\label{figure:albedo_g}
\end{figure}

In this work, we are primarily interested in measuring the phase function of P/La Sagra's nucleus for the purposes of deriving its physical size from its absolute magnitude (see below) and also being able to estimate the contribution due to dust to the comet's total observed flux during periods of activity (cf.\ Section~\ref{section:activedata}).  It would be useful of course, however, if we were also able to infer physical information from the slope parameter, $G$, as well.  Unfortunately, $G$ appears to have limited value as a diagnostic indicator of physical surface properties of individual asteroids.  There is substantial overlap in the distributions of $G$ values for asteroids classified as C-type and S-type asteroids, and several of the MBCs, including P/La Sagra, have $G$ parameter values within the range of values found for both taxonomic types (Figure~\ref{figure:taxonomy_g}).  As such, we are unable to infer taxonomic information for individual MBCs from their $G$ values alone.  Similarly, while there is a weak trend of increasing albedo with increasing $G$, as suggested by early studies \citep[e.g.,][]{har89}, there is also a substantial fraction of low-albedo asteroids for which a large range of $G$ values have been measured (Figure~\ref{figure:albedo_g}), indicating that $G$ is a poor indicator of the likely albedo of an object as well.  The distributions of $G$ values for different taxonomic types do peak at distinct values (cf.\ Figure~\ref{figure:taxonomy_g}), suggesting that a probabilistic analysis of the $G$ distribution of MBCs with respect to those of other asteroid taxonomic types could be useful once $G$ values are known for a substantially larger number of MBCs, even though $G$ is a poor diagnostic tool for individual objects \cite[e.g.,][]{osz12}.

We also fit a linear phase function to the inactive data, finding an absolute magnitude of $m(1,1,0)=18.7\pm0.2$~mag and a phase-darkening coefficient of $\beta=0.037\pm0.010$~mag~deg$^{-1}$.  This phase-darkening coefficient is comparable to values measured for other cometary nuclei including 10P/Tempel 2 \citep[$\beta=0.035\pm0.005$~mag~deg$^{-1}$;][]{jew89}, 28P/Neujmin 1 \citep[$\beta=0.025\pm0.006$~mag~deg$^{-1}$;][]{del01}, and 143P/Kowal-Mrkos \citep[$\beta=0.043\pm0.014$~mag~deg$^{-1}$;][]{jew03}, although several much larger phase-darkening coefficients have also been measured \citep[references in][]{sno11}.  It is also comparable to values measured for C-type asteroids \citep[$\beta_{mean}\approx 0.043$~mag~deg$^{-1}$; references in][]{bel00}.  The large $\beta$ is consistent with P/La Sagra having a low-albedo, possibly porous surface.
  
The effective nucleus radius (in km), $r_N$, of an object with an absolute magnitude of $H_R$ is given by
\begin{equation}
{p_{R}} r_{N}^2 = (2.24\times10^{16})\times 10^{0.4[m_{\odot}-H_R]}
\end{equation}
where $p_R$ is the geometric $R$-band albedo, and $m_{\odot}=-27.07$ is the apparent $R$-band magnitude of the Sun \citep{har80,har82,har90}.  Assuming a geometric $R$-band albedo of $p_R=0.05$, similar to that measured for other MBCs \citep{hsi09b}, we estimate an effective nucleus radius for P/La Sagra of $r_N=0.55\pm0.05$~km.

\subsection{Analysis of Active Data}
\label{section:activedata}

Using the phase function derived here, we can estimate the amounts of excess dust present in observations from 2010 and 2011 \citep[][this work]{hsi12c}.  The ratio of the total scattering surface area of dust, $A_d$, to that of the nucleus, $A_N$, is given by
\begin{equation}
{{A_d}\over{A_N}}={{1-10^{0.4(H_{R,tot} - H_R)}}\over{10^{0.4(H_{R,tot} - H_R)}}}
\label{equation:adan}
\end{equation}
where $H_{R,tot}$ is the equivalent absolute magnitude of the active nucleus at $R=\Delta=1$~AU and $\alpha=0^{\circ}$ computed using the $H,G$ phase function and the best-fit $G$ parameter determined above (assuming that the dust exhibits the same phase darkening behavior as the nucleus).  The equivalent ratio of the total mass of dust, $M_d$, to that of the nucleus, $M_N$, is then given by
\begin{equation}
{M_d\over M_N} = {{\bar a}\rho_d\over r_N\rho_N} \left({A_d\over A_N}\right)
\label{equation:mdmn}
\end{equation}
where ${\bar a}$ is the mean effective grain radius, $\rho_d$ is the dust grain density, and $\rho_N$ is the bulk density of the nucleus.  We can then compute the total dust scattering surface area using
\begin{equation}
A_d = \pi r_N^2 \left({A_d\over A_N}\right)
\label{equation:ad}
\end{equation}
and the total dust mass using
\begin{equation}
M_d = {4\over3}\pi r_N^2{\bar a}\rho_d \left({A_d\over A_N}\right)
\label{equation:md}
\end{equation}

\citet{mor11b} found particle radii for P/La Sagra ranging from $a_{\rm min}=5\times10^{-6}$~m to $a_{\rm max}=1\times10^{-2}$~m, assuming a power-law size distribution with an index of $q=3.5$.  Following \citet{jew14b}, we compute a mean effective particle radius (by mass), weighted by the size distribution, scattering cross-section, and residence time, using
\begin{equation}
{\bar a} = {{\int_{a_{\rm min}}^{a_{\rm max}} \! a \pi a^2 (Ka^{1/2})(N_0(R)a^{-q})da} \over {\int_{a_{\rm min}}^{a_{\rm max}} \! \pi a^2(Ka^{1/2})(N_0(R)a^{-q})da}} = {{\int_{a_{\rm min}}^{a_{\rm max}} \! a^{(3.5-q)}da} \over {\int_{a_{\rm min}}^{a_{\rm max}} \! a^{(2.5-q)}da}}
\end{equation}
where $K$ and $N_0(R)$ are constants related to the residence time of a particle of size $a$ and the reference dust production rate at a given heliocentric distance, $R$, respectively.
Using $q=3.5$ and assuming $a_{\rm max}\gg a_{\rm min}$, we then find
\begin{equation}
{\bar a} \sim {a_{\rm max}\over ln(a_{\rm max}/a_{\rm min})}
\end{equation}
which gives us ${\bar a}\sim 1$~mm using $a_{\rm min}$ and $a_{\rm max}$ from \citet{mor11b}. We use this effective mean particle radius to calculate the dust-to-nucleus scattering surface area ratios, dust-to-nucleus mass ratios, total dust scattering surface areas, and total dust masses from measurements reported by \citet{hsi12c} of the total flux of P/La Sagra.  These measurements were made using rectangular apertures enclosing the entire visible dust cloud (i.e., coma and tail) surrounding the comet in observations where a strong coma and a tail are visible (i.e., from 2010 to early 2011).  For observations where only a faint dust trail was visible (i.e., in late 2011), the trail's extremely low surface brightness and the abundance of bright nearby field stars prevented \citet{hsi12c} from performing reliable trail photometry, and so for those data, only the near-nucleus flux (i.e., from the nucleus and the unresolved coma) was measured.

Results are tabulated in Table~\ref{table:activedata}.  For these calculations, we assume $\rho_N=1400$~kg~m$^{-3}$ as determined to be characteristic of C-type asteroids \citep{bri02}, with which MBCs have been found to be similar \citep{lic11}.  For the dust grain density, we assume $\rho_d=2500$~kg~m$^{-3}$ as determined to be characteristic of CI and CM carbonaceous chondrite meteorites \citep{bri02}, which are generally believed to be associated with primitive, water-bearing C-type asteroids like the MBCs.

Following the analysis used in \citet{hsi12c}, we find that the total dust mass around P/La Sagra increases at a net rate of ${\dot M_d}\sim30$~kg~s$^{-1}$ between 2010 August and 2010 December.  This net dust production rate is two orders of magnitude higher than that computed by \citet{hsi12c} due to the assumption of larger mean effective grain sizes in this work.  After accounting for the higher dust grain density assumed here, this value is also a few times larger than the mass loss rate computed by \citet{mor11b}, not accounting for the effect of dust dissipation.  Given the strongly parameter-dependent and inherently underconstrained nature of dust modeling of comets in general, and the simplifications introduced here, however, we regard our results to be precise to an order of magnitude, at best.  As such, we find our results to be consistent with those of \citet{mor11b}.

As for the set of observations obtained between 2011 August and 2011 December where a residual dust trail was observed but no visible coma, we find a weighted average total dust mass of ${\bar M_d}\sim(1.1\pm0.6)\times10^5$~kg from the four photometric points from this time period.  This indicates that a low-level, unresolved coma, presumably consisting of large, slow-moving particles from the 2010 emission event that radiation pressure has not yet dispersed beyond the projected radius of the seeing disk on the sky, was likely present at the time.  We can therefore confirm that our omission of this data in our determination of P/La Sagra's phase function (Section~\ref{section:phsfn_analysis}) was indeed justified.

\setlength{\tabcolsep}{5.0pt}
\begin{table}[ht]
\caption{Analysis of 2010-2011 Photometry}
\smallskip
\footnotesize
\begin{tabular}{lccccccc}
\hline\hline
\multicolumn{1}{c}{UT Date}
 & \multicolumn{1}{c}{$\nu^a$}
 & \multicolumn{1}{c}{$m_{\rm tot}(R,\Delta,\alpha)^b$}
 & \multicolumn{1}{c}{$H_{R,tot}^c$}
 & \multicolumn{1}{c}{$A_d/A_N^d$}
 & \multicolumn{1}{c}{$M_d/M_N^e$}
 & \multicolumn{1}{c}{$A_d^f$}
 & \multicolumn{1}{c}{$M_d^g$} \\
\hline
2010 Jun 25     &   0.0 & \multicolumn{1}{c}{\it Perihelion} & ... & ... & ... & ... & ... \\
2010 Aug 16     &  12.9 & $18.3\pm0.3$ & $14.1\pm0.3$ &   $50\pm15$ & $(17\pm5)\times 10^{-5}$  &  $(5\pm1)\times 10^7$ &  $(15\pm4)\times10^7$   \\
2010 Sep 08     &  18.5 & $18.0\pm0.3$ & $14.0\pm0.3$ &   $60\pm20$ & $(19\pm6)\times 10^{-5}$  &  $(5\pm2)\times 10^7$ &  $(18\pm5)\times10^7$   \\
2010 Sep 19 (1) &  21.3 & $17.9\pm0.3$ & $13.8\pm0.3$ &   $65\pm20$ & $(22\pm6)\times 10^{-5}$  &  $(6\pm2)\times 10^7$ & $(20\pm6)\times10^7$   \\
2010 Sep 19 (2) &  21.3 & $17.8\pm0.3$ & $13.7\pm0.3$ &   $70\pm20$ & $(24\pm7)\times 10^{-5}$  &  $(7\pm2)\times 10^7$ & $(22\pm6)\times10^7$   \\
2010 Sep 29     &  23.7 & $18.0\pm0.3$ & $13.9\pm0.3$ &   $65\pm20$ & $(21\pm6)\times 10^{-5}$  &  $(6\pm2)\times 10^7$ & $(20\pm6)\times10^7$   \\
2010 Oct 05     &  25.2 & $18.0\pm0.3$ & $13.8\pm0.3$ &   $70\pm20$ & $(23\pm7)\times 10^{-5}$  &  $(6\pm2)\times 10^7$ & $(21\pm6)\times10^7$   \\
2010 Oct 19     &  28.5 & $18.1\pm0.3$ & $13.7\pm0.3$ &   $75\pm20$ & $(25\pm7)\times 10^{-5}$  &  $(7\pm2)\times 10^7$ & $(23\pm7)\times10^7$   \\
2010 Nov 26     &  37.8 & $18.4\pm0.3$ & $13.4\pm0.3$ &  $100\pm30$ & $(33\pm10)\times 10^{-5}$ &  $(9\pm3)\times 10^7$ & $(30\pm9)\times10^7$   \\
2010 Dec 12     &  41.3 & $18.4\pm0.3$ & $13.2\pm0.3$ &  $120\pm35$ & $(40\pm11)\times 10^{-5}$ & $(11\pm3)\times 10^7$ & $(36\pm10)\times10^7$   \\
2010 Dec 31     &  45.9 & $18.2\pm0.3$ & $12.8\pm0.3$ &  $175\pm50$ & $(57\pm16)\times 10^{-5}$ & $(16\pm5)\times 10^7$ & $(53\pm15)\times10^7$   \\
2011 Feb 03     &  53.5 & $19.1\pm0.3$ & $13.5\pm0.3$ &   $90\pm30$ & $(31\pm9)\times 10^{-5}$  &  $(8\pm2)\times 10^7$ & $(28\pm8)\times10^7$   \\
2011 Aug 26     &  95.1 & $24.1\pm0.3$ & $18.2\pm0.3$ & $0.2\pm0.3$ & $(8\pm12)\times 10^{-7}$  &  $(2\pm3)\times 10^5$ & $(7\pm11)\times10^5$    \\
2011 Aug 31     &  96.0 & $23.4\pm0.3$ & $17.5\pm0.3$ & $1.3\pm0.6$ & $(43\pm21)\times 10^{-7}$ & $(12\pm6)\times 10^5$ & $(39\pm20)\times10^5$  \\
2011 Sep 25     & 100.5 & $23.8\pm0.3$ & $18.1\pm0.3$ & $0.4\pm0.4$ & $(12\pm13)\times 10^{-7}$ &  $(3\pm4)\times 10^5$ & $(11\pm12)\times10^5$   \\
2011 Dec 31     & 117.0 & $23.0\pm0.3$ & $18.2\pm0.3$ & $0.2\pm0.3$ & $(6\pm11)\times 10^{-7}$  &  $(2\pm3)\times 10^5$ & $(6\pm10)\times10^5$   \\
2013 Mar 12     & 180.0 & \multicolumn{1}{c}{\it Aphelion}   & ... & ... & ... & ... & ... \\
2015 Nov 30     &   0.0 & \multicolumn{1}{c}{\it Perihelion} & ... & ... & ... & ... & ... \\
\hline
\hline
\end{tabular}
\newline {$^a$ True anomaly, in degrees}
\newline {$^b$ Total observed apparent $R$-band magnitude, where uncertainties are assumed to be dominated by the unknown rotational phase of the nucleus, and where we assume $\Delta m = 0.6$~mag}
\newline {$^c$ Equivalent absolute $R$-band magnitude at $R=\Delta=1$~AU and $\alpha=0^{\circ}$}
\newline {$^d$ Ratio of total scattering surface area of dust to that of the nucleus}
\newline {$^e$ Estimated dust-to-nucleus mass ratio, assuming $\rho_N=1400$~kg~m$^{-3}$, $\rho_d=2500$~kg~m$^{-3}$, and ${\bar a}=1$~mm}
\newline {$^f$ Estimated total scattering surface area of dust grains, in m$^2$}
\newline {$^g$ Estimated total dust mass, in kg, assuming $\rho_d=2500$~kg~m$^{-3}$}
\label{table:activedata}
\end{table}

\section{DISCUSSION}
\label{section:discussion}

\subsection{Comparison to Other MBCs}
\label{section:mbc_comparison}

\begin{figure}
\centerline{\includegraphics[width=6.0in]{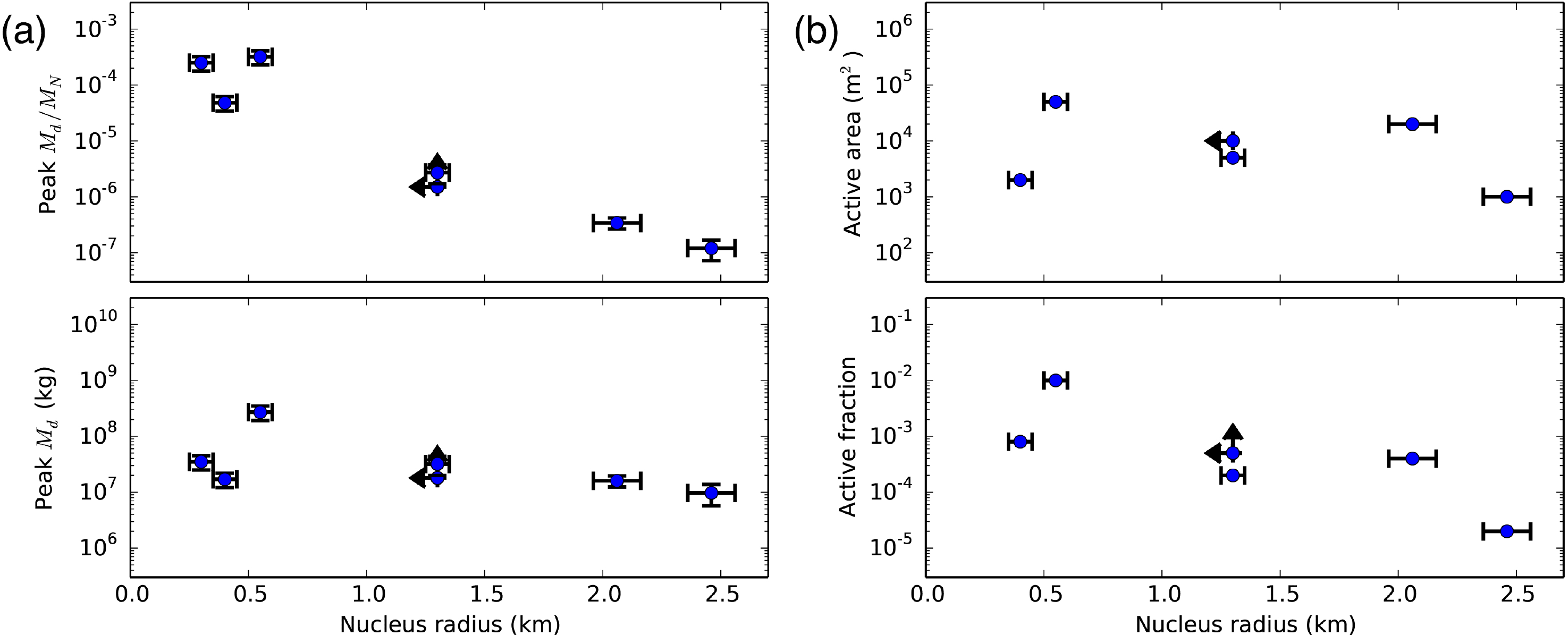}}
\caption{\small (a) Plots of the peak dust-to-nucleus mass ratios (top panel) and the peak total dust masses (bottom panel) measured for the known MBCs as functions of nucleus size (cf.\ Table~\ref{table:mbcdata}). (b) Plots of the total active surface areas (top panel) and the active surface fractions (bottom panel) inferred for the MBCs.  In all panels, points with ordinary error bars denote MBCs with well-constrained nucleus radii (133P, 176P, 238P, 259P, P/La Sagra, and P/2006 VW$_{139}$), and points with arrows instead of error bars (representing P/2012 T1) indicate that only an upper-limit nucleus radius (and therefore lower-limit dust-to-nucleus mass ratios, total dust masses, and active surface fractions) is known.
}
\label{figure:size_vs_activity}
\end{figure}

Having computed revised measurements of P/La Sagra's nucleus size and activity strength, we now wish to compare the comet to other MBCs (cf.\ Section~\ref{section:intro_lasagra}).  We use previously reported photometry of active dust emission and computed absolute magnitudes of inactive MBC nuclei (references in Table~\ref{table:mbcdata}) and Equations~\ref{equation:adan}-\ref{equation:md} to compute the peak dust-to-nucleus mass ratios ($M_d/M_N$) and peak total inferred dust masses ($M_d$) associated with seven of the eight MBCs known to date (Table~\ref{table:mbcdata}).  For consistency, we assume mean effective grain sizes of ${\bar a}=1$~mm, bulk nucleus densities of $\rho_N=1400$~kg~m$^{-3}$, and dust grain densities of $\rho_d=2500$~kg~m$^{-3}$ in all of these calculations.  We then plot these derived quantities as functions of nucleus radius, which are all computed using previously reported absolute magnitudes from the literature and assumed albedos of $p_R=0.05$ (Figure~\ref{figure:size_vs_activity}a).  In the case of P/2012 T1 (PANSTARRS), only a lower-limit absolute magnitude is currently available, and so only an upper-limit nucleus size and lower-limit peak $M_d/M_N$ and $M_d$ values are reported here.  Meanwhile, we omit P/2013 R3 (Catalina-PANSTARRS) from consideration in this analysis due to the clearly unusual physical circumstances that led to the object's complete disintegration, and the therefore low likelihood that any comparisons to other MBCs will be particularly physically meaningful.

\setlength{\tabcolsep}{3pt}
\begin{table}[ht]
\caption{Main-Belt Comet Activity}
\smallskip
\scriptsize
\begin{tabular}{lcccccccccc}
\hline\hline
 \multicolumn{1}{c}{Name} &
 \multicolumn{1}{c}{$q^a$} &
 \multicolumn{1}{c}{$H_R^b$} &
 \multicolumn{1}{c}{$r_N^c$} &
 \multicolumn{1}{c}{${\dot m_w}(q)^d$} &
 \multicolumn{1}{c}{${\dot M_d}^e$} &
 \multicolumn{1}{c}{$A_{\rm act}^f$} &
 \multicolumn{1}{c}{$f_{\rm act}^g$} &
 \multicolumn{1}{c}{${M_d/M_N}^h$} &
 \multicolumn{1}{c}{$M_d^i$} &
 \multicolumn{1}{c}{Refs.$^j$} \\ 
\hline
133P/Elst-Pizarro & 2.650 & 15.49$\pm$0.05 & 2.06$\pm$0.05 & $7.2\times10^{-6}$ & 1.4 & $2\times10^4$ & $4\times10^{-4}$  & (6.1$\pm$1.4)$\times$10$^{-7}$ & $(3.1\pm0.7)\times10^7$ & [1]  \\
176P/LINEAR       & 2.577 & 15.10$\pm$0.05 & 2.46$\pm$0.05 & $8.3\times10^{-6}$ & 0.1 & $1\times10^3$ & $2\times10^{-5}$  & (2.1$\pm$0.9)$\times$10$^{-7}$ & $(1.9\pm0.8)\times10^7$ & [2]  \\
238P/Read         & 2.365 & 19.05$\pm$0.05 & 0.40$\pm$0.05 & $1.3\times10^{-5}$ & 0.2 & $2\times10^3$ & $8\times10^{-4}$  & (8.6$\pm$2.5)$\times$10$^{-5}$ & $(3.2\pm1.0)\times10^7$ & [3]  \\
259P/Garradd      & 1.793 & 19.71$\pm$0.05 & 0.29$\pm$0.05 & $3.4\times10^{-5}$ & --- & ---           & ---               & (4.5$\pm$1.3)$\times$10$^{-4}$ & $(6.8\pm1.9)\times10^7$ & [4]  \\
P/2006 VW$_{139}$ & 2.434 & 16.5$\pm$0.1   & 1.29$\pm$0.05 & $1.1\times10^{-5}$ & 0.5 & $5\times10^3$ & $2\times10^{-4}$  & (4.8$\pm$1.8)$\times$10$^{-6}$ & $(6.1\pm2.2)\times10^7$ & [5]  \\
P/2010 R2         & 2.622 & 18.4$\pm$0.2   & 0.54$\pm$0.05 & $7.6\times10^{-6}$ &  4  & $5\times10^4$ & $1\times10^{-2}$  & (5.7$\pm$1.6)$\times$10$^{-4}$ & $(5.3\pm1.5)\times10^8$ & [6]  \\
P/2012 T1         & 2.411 & $>$16.5        & $<$1.3        & $1.1\times10^{-5}$ & 1.2 & $1\times10^4$ & $>5\times10^{-4}$ & $>$2.7$\times$10$^{-6}$        & $>$3.4$\times$10$^7$    & [7]  \\
P/2013 R3         & 2.204 & $>$15.1        & $<$2.5        & $1.7\times10^{-5}$ & --- & ---           & ---               & ---                            & ---                     & [8]  \\
\hline
\hline
\end{tabular}
\newline {$^a$ Osculating perihelion distance, in AU}
\newline {$^b$ Absolute $R$-band magnitude}
\newline {$^c$ Effective nucleus radius, in km, derived from $H_R$ and assuming $p_R=0.05$}
\newline {$^d$ Specific mass loss rate of water ice due to sublimation at perihelion, in kg~s$^{-1}$}
\newline {$^e$ Reported dust mass loss rates derived from numerical dust modeling (see References), in kg~s$^{-1}$} 
\newline {$^f$ Inferred active surface area, in m$^2$, assuming $f_{dg}=10$}
\newline {$^g$ Inferred active surface fraction}
\newline {$^h$ Estimated dust-to-nucleus mass ratio, assuming $\rho_N=1400$~kg~m$^{-3}$, $\rho_d=2500$~kg~m$^{-3}$, and ${\bar a}=1$~mm}
\newline {$^i$ Estimated total dust mass, in kg, assuming $\rho_d=2500$~kg~m$^{-3}$}
\newline {$^j$ References:
[1] \citet{hsi04,hsi09b,hsi10b,jew14b};
[2] \citet{hsi09b,hsi11a};
[3] \citet{hsi09a,hsi11b};
[4] \citet{jew09,mac12};
[5] \citet{hsi12b,lic13}; Hsieh et al., in prep;
[6] \citet{mor11b,hsi12c};
[7] \citet{hsi13a,mor13};
[8] \citet{jew14a}
}
\label{table:mbcdata}
\end{table}

Intriguingly, there is a clear trend of increasing peak $M_d/M_N$ values with decreasing nucleus radius (Figure~\ref{figure:size_vs_activity}a, upper panel).  The cause of this trend is evident from the lower panel of Figure~\ref{figure:size_vs_activity}a, where we can see that while the minimum and maximum $M_d/M_N$ values measured for MBCs differ by over three orders of magnitude, $M_d$ values vary by less than two orders of magnitude, or if a single outlier (P/La Sagra) is removed from consideration, less than one order of magnitude.  As such, the strong anti-correlation of $M_d/M_N$ with nucleus size is due nearly entirely to the fact that almost all of the MBCs appear to eject nearly identical quantities of dust, despite varying by nearly three orders of magnitude in volume (and therefore in inferred mass).

This consistency among the inferred total dust masses measured for the first five known MBCs was first noted in \citet{hsi12c}, and so it is interesting that it appears to remain viable for the present sample of seven of the eight currently known MBCs.  As noted by \citet{hsi12c}, however, the interpretation of this trend is not straightforward.  The total dust mass observed for each object is essentially the net dust production rate (i.e., the difference between the mass loss rate of the nucleus and the rate at which the ejected dust is dispersed by radiation pressure to the point of non-detectability) integrated over the period starting from the start of dust emission until the time of observation, where in principle, the mass loss rate of the nucleus is related to the effective size of the active area on the MBC (i.e., the area of exposed volatile material available for sublimation).  However, a large mass loss rate could be coupled with a rapid dust dissipation rate to produce the same net dust production rate (and therefore the same total observed dust mass) as an object with a smaller mass loss rate coupled with a slower dust dissipation rate.  Since the known MBCs are all observed to be active at similar heliocentric distances, one might expect dissipation rates to also be similar. Other factors can also affect dust observability, however.  For instance, dust confined to an object's orbit plane will have a greater optical depth when viewed edge-on than when viewed at a non-zero angle to the orbit plane.  Therefore, dust emitted from a low-inclination MBC (such as 133P or 176P; $i=1.4^{\circ}$ and $i=0.2^{\circ}$, respectively) should remain observable for a longer period of time than dust from a high-inclination MBC (such as P/La Sagra or 259P; $i=21.4^{\circ}$ and $i=15.9^{\circ}$, respectively).

Given the numerous factors that can influence the observed appearance of dust emission, numerical dust modeling is typically employed to infer properties of interest, such as ejection velocities and mass loss rates.  The determination of the dust production rate of a MBC in this manner, though, is highly parameter-dependent, where final results can vary widely due to different assumptions of parameters such as the optically dominant grain size and bulk densities of nucleus and dust material.  Converting dust production rates to sublimation rates and then to an active surface area is also complicated by such factors as the typical lack of observational constraints on gas-to-dust ratios and the thermal parameters of the surface material in question, the possible action of jets \citep[cf.][]{yel04,hsi11a,bel13,far13}, and the time-dependent activity-quenching effects of mantling \citep[cf.][]{jew96}.  Finally, it is possible that the apparent trend of similar dust masses is an observational selection effect, and that MBCs exhibiting smaller total dust masses are simply not easily discovered by currently available facilities (even though lower levels of activity can be observed once an object is known to be active and is therefore specifically targeted for monitoring).

Despite the many caveats that must be considered, for reference, we follow the procedure used by \citet{hsi12c} to compute the area-dependent sublimation rates of surface ice at perihelion for the known MBCs, and then compare them to available reported mass loss rates \citep{hsi09a,hsi11a,mor11b,mor13,jew14b}, to compute equivalent active areas and active fractions (Table~\ref{table:mbcdata}).  Finally, we plot these values as functions of nucleus radius in Figure~\ref{figure:size_vs_activity}b.  For these calculations, we use the following energy balance equation for the equilibrium temperature, $T$, of a sublimating surface:
\begin{equation}
{F_{\odot}\over R^2}(1-A) = \chi\left[\varepsilon\sigma T^4 + L{\dot m_w}(T)\right]
\end{equation}
where $F_{\odot}=1360$~W~m$^{-2}$ is the solar constant, $R$ is the heliocentric distance of the object in AU, $A=0.05$ is the assumed Bond albedo of the body, $\varepsilon=0.9$ is the assumed effective infrared emissivity, $\sigma$ is the Stefan-Boltzmann constant, $L=2.83$~MJ~kg$^{-1}$ is the latent heat of sublimation for water ice, ${\dot m_w}$ is the temperature-dependent mass loss rate of water ice due to sublimation, and $\chi$ describes the distribution of solar heating over an object's surface, assumed here to be $\chi=2.5$ (between the so-called subsolar and isothermal extremes of $\chi=1$ and $\chi=4$, respectively).  We then compute the active surface area needed to produce the model-determined mass loss rates using
\begin{equation}
A_{\rm act} = {{\dot M_d}\over {\dot m_w}f_{dg}}
\end{equation}
where $f_{dg}$ is the dust-to-gas production rate ratio (by mass).  Dust-to-gas ratios of $f_{dg}=1$ are often assumed for comets, but as discussed in \citet{jew14b}, values of $f_{dg}>1$ are physically possible when ejected dust carrying more mass than escaping gas travels at much slower velocities than the gas, thus conserving momentum.  Like P/La Sagra, 2P/Encke has been observed to exhibit dust emission determined to be dominated by mm-sized grains, and ultraviolet and infrared observations have constrained its dust-to-gas ratio to $10\lesssim f_{dg}\lesssim30$ \citep{rea00}.  For our analysis here, we assume $f_{dg}=10$ \citep[cf.][]{jew14b}.

Finally, we compute the active surface fraction using
\begin{equation}
f_{\rm act} = {A_{\rm act}\over 4\pi r_N^2}
\end{equation}

We find that all of the inferred active surface areas computed here are within two orders of magnitude of each other, suggesting that if impact activation is indeed responsible for MBC activity \citep[cf.][]{hsi04,cap12}, the active areas on all of the MBCs considered here could have been excavated by similarly sized impactors.  Meanwhile, we note that 133P, 238P, P/2006 VW$_{139}$, and P/2012 T1 have similar inferred active surface fractions within an order of magnitude, while P/La Sagra's active fraction is about an order of magnitude larger than those of the other MBCs, and 176P's active fraction is roughly an order of magnitude smaller.  
The active surface areas and active surface fractions inferred here are also all at least an order of magnitude smaller than those of the vast majority of classical comets \citep[e.g.,][]{ahe95,fer99,sam13}, although MBC perihelion distances are also all generally larger than those of most Jupiter family and long-period comets that are observed.  
Considering the numerous caveats detailed above, we consider a more detailed investigation of the particular implications of the inferred active areas and active fractions of MBCs to be premature, but suggest that continued attention to inferred active surface areas as more MBCs are discovered and physically characterized could provide valuable insights into the origin and evolution of MBC activity.

\subsection{Future Observations}
\label{section:future_observations}

As discussed above (Section~\ref{section:background}), the detection of recurrent activity for a main-belt object during subsequent orbit passages is currently considered one of the most reliable indicators of sublimation-driven activity \citep{hsi12a,jew12}.  P/La Sagra will make its first return to perihelion since its 2010 discovery on 2015 November 30, and so monitoring of the object around that time will be extremely important for clarifying the nature of its activity.  Fortunately, unlike 259P, which had limited visibility during its first post-discovery perihelion passage in 2013 January 25 \citep{mac12}, P/La Sagra will be well-placed for observations to search for recurrent activity during its upcoming perihelion passage.

\begin{figure}
\centerline{\includegraphics[width=4.3in]{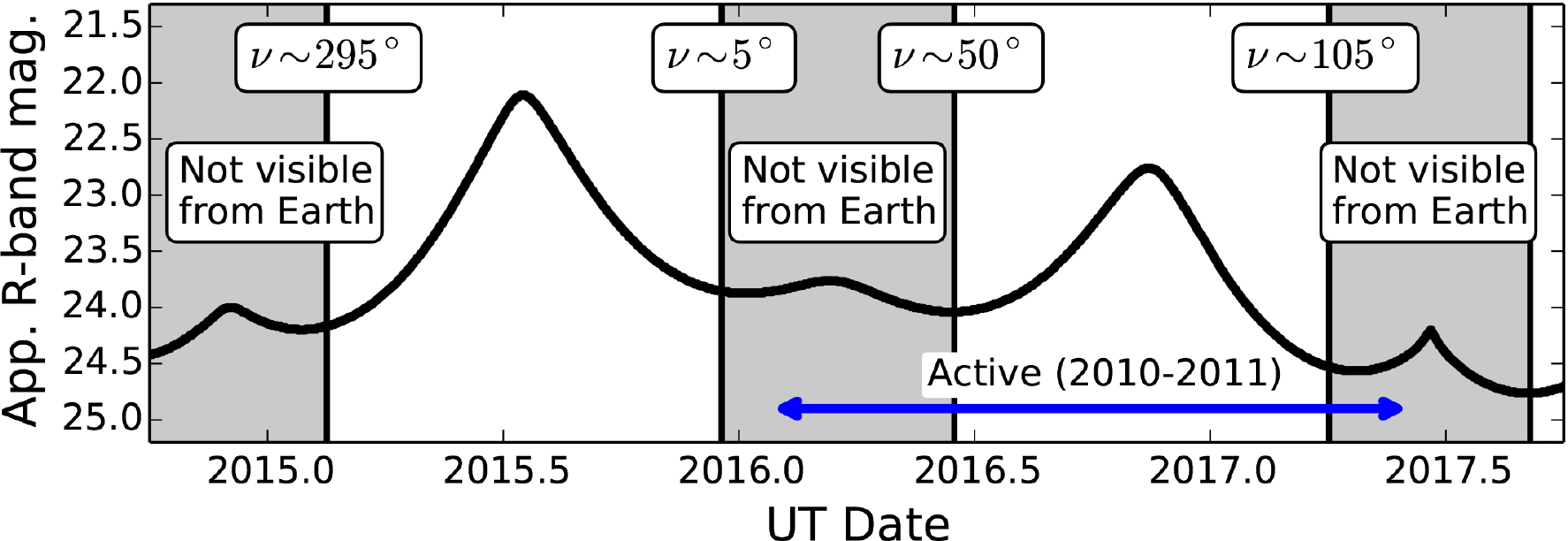}}
\caption{\small Plot of predicted apparent $R$-band magnitudes for the nucleus of P/La Sagra in the absence of activity from 2014 Oct 1 through 2017 Oct 1, a period which spans the object's 2015 perihelion passage.  Shaded regions mark time periods during which the comet is not observable from the Earth (i.e., solar elongation $<50^{\circ}$).  The true anomalies of the comet at the beginning and end of each observing window are labeled, while the true anomaly range over which the comet was previously observed to be active in 2010 and 2011 is marked with a blue double-headed arrow.
}
\label{figure:magnitude_predictions}
\end{figure}

We plot P/La Sagra's predicted magnitudes and available upcoming observing windows in Figure~\ref{figure:magnitude_predictions}.  P/La Sagra will be observable from the Southern hemisphere from 2015 February to 2015 November over a true anomaly range of $295^{\circ}\lesssim\nu\lesssim355^{\circ}$, and will be observable from the Northern hemisphere from 2015 March to 2015 December, or from $\nu\sim300^{\circ}$ to $\nu\sim5^{\circ}$.  It then will be observable from the North again from 2016 June to 2017 March over a true anomaly range of $50^{\circ}\lesssim\nu\lesssim105^{\circ}$. Due to its extremely northerly declination ($\delta>25^{\circ}$) in during the 2016-2017 observing window, however, it will not be well-placed for observing from the South during this time period.

For reference, we also mark the true anomaly range over which the comet was previously observed to be active during its 2010-2011 active period in Figure~\ref{figure:magnitude_predictions}.  We note though that we do not expect activity to be limited to that range.  P/La Sagra was observed to be active as early as $\nu=12.9^{\circ}$ during its 2010 active period \citep{hsi12c}, which it will reach again on 2016 January 20.  However, considering that the comet was already strongly active at the time of its discovery, and that 133P and 238P have both been observed to exhibit pre-perihelion activity, we expect that a much earlier turn-on point for P/La Sagra could be possible.  \citet{bau12} reported that {\it Wide-field Infrared Survey Explorer} observations of P/La Sagra from 2010 June 9-11, when the object was at $\nu\sim355^{\circ}$, showed no clear evidence of activity, but this non-detection could be attributable to insufficient image sensitivity.  238P was observed to be active as early as $\nu=306.1^{\circ}$ \citep{hsi11b}, which P/La Sagra will reach on 2015 April 19.  As such, monitoring P/La Sagra starting at least this early would be highly desirable.  In that regard, the nucleus size determination for the comet reported here will be extremely useful, as it will enable early detection of activity to be accomplished via photometric analysis even if resolved activity is not immediately observable.

Some deviations (which should average to zero for a sufficiently large data set) from the magnitude predictions plotted in Figure~\ref{figure:magnitude_predictions} are expected due to rotational brightness variations of the nucleus.  However, photometric measurements made for the comet that are consistently brighter than expected will represent strong evidence for the presence of unresolved activity.  Analogous analyses were used to discover the activity of 95P/(2060) Chiron \citep{bus88,tho88,mee89,har90}, detect the reactivation of 67P/Churyumov-Gerasimenko in 2007 \citep{sno13}, and to confirm the absence of recurrent activity for 176P in 2011 \citep{hsi14}.  If resolved cometary activity is immediately observable, use of this technique of using photometry to detect activity will of course be unnecessary.  In this case, however, it will still be useful to monitor P/La Sagra through both observing windows shown in Figure~\ref{figure:magnitude_predictions} and determine the amount of dust present at any given time (using the nucleus size measured in this work).  Doing so will enable us to monitor changes over time and also to enable comparisons to be made between P/La Sagra's activity strength over successive orbit passages, which will in turn allow us to better understand the rate of MBC devolatization and set limits on their active lifetimes.

\section{SUMMARY AND CONCLUSIONS}

We present recent observations of main-belt comet P/2010 R2 (La Sagra) obtained in 2011 and 2013 using the Gemini North telescope, and report the following key results:
\begin{enumerate}
\item{Following its 2010 September discovery, P/La Sagra continued to exhibit a dust trail, likely composed of large dust particles remaining from the 2010 active period, as late as 2011 December 31, when the comet was at $\nu=117^{\circ}$.  No activity was observed by the time of our next observations on 2013 March 3, when the comet was at $\nu=178.8^{\circ}$, or just before aphelion.  Using only photometry measured when the comet was observed to be inactive, we find best-fit IAU phase function parameters of $H_R=18.4\pm0.2$~mag and $G=0.17\pm0.10$, corresponding to an effective nucleus radius of $r_N=0.55\pm0.05$~km (assuming $p_R=0.05$).}
\item{We revisit photometry obtained when P/La Sagra was observed to be active in 2010 using our revised determination of the object's nucleus size, and find a peak dust-to-nucleus mass ratio of $M_d/M_N = (5.8\pm1.6)\times10^{-4}$, corresponding to an estimated total peak dust mass of $M_d = (5.3\pm1.5)\times10^8$~kg (assuming ${\bar a}=1$~mm, $\rho_N=1400$~kg~m$^{-3}$, and $\rho_d=2500$~kg~m$^{-3}$).  We also compute the inferred peak total active surface area and active surface fraction of P/La Sagra, finding $A_{\rm act}\sim5\times10^4$~m$^2$ and $f_{\rm act} \sim0.01$, respectively.}
\item{We confirm that the finding of \citet{hsi12c} that the total peak dust masses observed for MBCs are largely similar (within an order of magnitude), regardless of nucleus size, remains valid for seven MBCs out of the currently known population of eight for which physical characterizations are available.  Inferred active surface areas derived from mass loss rates determined from numerical dust modeling are also found to be similar (within two orders of magnitude).}
\item{We discuss P/La Sagra's upcoming perihelion passage in 2015, its first since its discovery in 2010, particularly focusing on the opportunities available for follow-up observations.  The comet will be observable from the Earth from 2015 February through December, during which it will cover a true anomaly range from $\nu\sim295^{\circ}$ to $\nu\sim5^{\circ}$, and then from 2016 June through 2017 March, during which it will cover a true anomaly range of $50^{\circ}\lesssim\nu\lesssim105^{\circ}$.  Observations during these periods are highly encouraged, as they will provide excellent opportunities for searching for recurrent activity, which would be strong evidence that P/La Sagra's activity is in fact sublimation-driven, as well as to obtain data enabling direct comparison of P/La Sagra's activity strength over successive orbit passages, where information enabling the photometric detection of unresolved activity, if needed, is provided by this work.}
\end{enumerate}
In addition to efforts to discover more MBCs \citep[e.g.,][]{gil09,son11,was13}, continued efforts to physically characterize known MBCs well after their discoveries should be considered a high priority.  Studying the nuclei of currently known MBCs will improve our understanding of their activity, as shown in this work, as well as provide insights into how the active MBC population relates to the inactive background asteroid population, e.g., by determining how their size distributions compare to one another.  Given the scarcity of known MBCs, it is imperative that we learn as much as we can about each of the few objects that are known in order to begin to discern the characteristics of the much larger, but mostly unknown, total population.  Such work should give us valuable insights into the nature and origin of this population, and by extension, into volatile abundance, preservation, and evolution in the inner solar system in general.

\section*{Acknowledgements}
We thank David Jewitt, Bin Yang, Yan Fern\'andez, and two anonymous referees for helpful comments on this manuscript.
Support for this work was provided by NASA to HHH through Hubble Fellowship grant HF-51274.01 awarded by the Space Telescope Science Institute, which is operated by the Association of Universities for Research in Astronomy, Inc., for the National Aeronautics and Space Administration (NASA), under contract NAS 5-26555.
Observations were obtained at the Gemini Observatory (observing programs GN-2011B-Q-17 and GN-2013A-Q-102), which is operated by the Association of Universities for Research in Astronomy, Inc., under a cooperative agreement with the NSF on behalf of the Gemini partnership: the National Science Foundation (United States), the National Research Council (Canada), CONICYT (Chile), the Australian Research Council (Australia), Minist\'{e}rio da Ci\^{e}ncia, Tecnologia e Inova\c{c}\~{a}o (Brazil) and Ministerio de Ciencia, Tecnolog\'{i}a e Innovaci\'{o}n Productiva (Argentina), where we thank J.\ Chavez, D.\ Coulson, P.\ Hirst, M.\ Hoenig, J.\ Kemp, E.\ Martioli, R.\ Mason, A.\ Smith, A.\ Stephens, S.\ Stewart, and B.\ Walp for their assistance in obtaining these observations.
The Pan-STARRS1 Surveys (PS1), which provided calibration data for this work, have been made possible through contributions of the Institute for Astronomy, the University of Hawaii, the Pan-STARRS Project Office, the Max-Planck Society and its participating institutes, the Max Planck Institute for Astronomy, Heidelberg and the Max Planck Institute for Extraterrestrial Physics, Garching, The Johns Hopkins University, Durham University, the University of Edinburgh, Queen's University Belfast, the Harvard-Smithsonian Center for Astrophysics, the Las Cumbres Observatory Global Telescope Network Incorporated, the National Central University of Taiwan, the Space Telescope Science Institute, the National Aeronautics and Space Administration under Grant No.\ NNX08AR22G issued through the Planetary Science Division of the NASA Science Mission Directorate, the National Science Foundation under Grant No.\ AST-1238877, the University of Maryland, and Eotvos Lorand University (ELTE).  We thank the PS1 Builders and PS1 operations staff for construction and operation of the PS1 system and access to the data products provided.  This publication makes use of data products from the Wide-field Infrared Survey Explorer, which is a joint project of the University of California, Los Angeles, and the Jet Propulsion Laboratory (JPL)/California Institute of Technology, funded by NASA, and NEOWISE, which is a project of JPL/California Institute of Technology, funded by the Planetary Science Division of NASA.

\end{document}